# Single-electron transport in a molecular Hubbard dimer

*James O. Thomas[a,b]\*, Jakub K. Sowa[c], Bart Limburg[a,b], Xinya Bian[a], Charalambos Evangeli[a], Jacob L. Swett[a], Sumit Tewari[a], Jonathan Baugh[d], George C. Schatz[c], G. Andrew D. Briggs[a], Harry L. Anderson[b] and Jan A. Mol[e]*

[a] Department of Materials, University of Oxford, Parks Road, Oxford, OX1 3PH, UK

[b] Department of Chemistry, University of Oxford, Chemistry Research Laboratory, Oxford OX1 3TA, UK

[c] Department of Chemistry, Northwestern University, Evanston, Illinois 60208, USA

[d] Institute for Quantum Computing, University of Waterloo, Waterloo, ON, N2L 3G1, Canada

[e] School of Physics and Astronomy, Queen Mary University of London, London, E1 4NS, UK

\* Email: james.thomas@materials.ox.ac.uk

**Many-body electron interactions are at the heart of chemistry and solid-state physics. Understanding these interactions is crucial for the development of molecular-scale quantum and nanoelectronic devices. Here, we investigate single-electron tunneling through an edge-fused porphyrin oligomer and demonstrate that its transport behavior is well described by the Hubbard dimer model. This allows us to study the role of electron-electron interactions in the transport setting. In particular, we empirically determine the molecule's on-site and inter-site electron-electron repulsion energies, which are in good agreement with density functional calculations, and establish the molecular electronic structure within the various charge states. The gate-dependent rectification behavior is used to further confirm the selection rules and state degeneracies resulting from the Hubbard model. We therefore demonstrate that current flow through the molecule is governed by a non-trivial set of vibrationally coupled electronic transitions between various many-body states, and experimentally confirm the importance of electron-electron interactions in single-molecule devices.**

Charge transport is one of the key observables in quantum systems, yet its interpretation is often complicated by strong many-body correlations. In molecular systems, these electron-electron and electron-vibration interactions are especially important in the resonant transport regime, and a rich tapestry of transport and out-of-equilibrium phenomena has been observed in single-molecule junctions.[1-7] For most single-molecule junctions these phenomena are limited to local interactions, including the observation of Coulomb blockade (and related Pauli blockade) and Franck-Condon blockade. In extended molecular systems, more intricate interacting approaches such as the fermionic Hubbard model that account for electron-electron interactions beyond the observation of Coulomb blockade[8-15] have been shown to be important in describing experimental results[16-18].

The Hubbard model is a ubiquitous description of strongly correlated condensed matter systems, including high-temperature superconductors and topological insulators. From a molecular perspective, the fermionic Hubbard model is an extension to the non-interacting Hückel model, which has been used very successfully in combination with Landauer theory to describe off-resonance quantum transport through extended molecules, but fails in the resonant transport regime where interactions become dominant.[19] By contrast, the Hubbard model not only considers the kinetic 'hopping' terms but also accounts for the Coulomb potentials, making it an extremely useful tool to empirically parameterize the many-body interactions that make up molecular structure-property relations.

Here, we investigate charge transport through an edge-fused porphyrin trimer, **FP3**, (Fig. 1a) that is weakly coupled to the source and drain electrodes through two electron-rich pyrene anchor groups. Unlike in most single-molecule junction devices where only one or two charge-states are accessible,[20, 21] the highly redox-active properties of this fully conjugated oligomer enable us to study up to four charge-states. This in turn lets us measure the addition energies and out-of-equilibrium current rectification that are a result of electron transfers between the many-body Fock states which arise from partial filling of the two highest occupied molecular orbitals. As these orbitals have high electron density on the two end groups of the molecule, we can interpret the results in the framework of a Hubbard dimer, the simplest non-trivial Hubbard Hamiltonian, to quantify the strength of the electron-electron and electron-vibration interactions and determine the Dyson coefficients that correspond to the wavefunction overlap between the Fock states.

**Results and Discussion**

**Molecular Devices**

We have designed the molecule **FP3** such that it contains two electron-rich anchoring groups separated by a conjugated edge-fused porphyrin trimer.[22] The three bonds between each porphyrin result in a planar structure and thus enhance electron delocalization across **FP3**.[23] The electrochemical gap of **FP3** from square-wave voltammetry is 0.8 eV (as compared to 1.7 eV for a zinc porphyrin monomer with the same anchor groups[20]). The longest wavelength absorbance maximum in the optical absorbance spectrum of **FP3** at 1500 nm (0.83 eV), compared to 700 nm (1.77 eV) for the monomer (**FP3** spectra are in the SI).

The single-molecule device architecture is shown in Fig. 1b and is described in more detail in the Methods Section. Briefly, graphene source and drain electrodes, separated by approximately 1-2 nm, are fabricated by electron-beam lithography and feedback-controlled electroburning.[24, 25] A solution (2 µM in toluene) of **FP3** is drop-cast on the electrodes. The tridodecyloxypyrene (TDP) anchor groups on the periphery of the fused porphyrin unit interact with the graphene electrodes through a π-stacking interaction that leads to weak molecule-electrode coupling,[20] while the aryl groups (Ar, Fig. 1a) prevent molecular aggregation (see SI). The gate electrode is either the doped silicon substrate with a thermally grown 300 nm $SiO_2$ dielectric (device **A**, device **B**) or gold with a 10 nm layer of $HfO_2$ dielectric grown by atomic-layer deposition (device **C**, and shown in the Fig. 1b). Stability diagrams prior to molecular deposition are included in the SI and confirm the signals observed are due to the deposition of **FP3**, and not residual carbon quantum dots from the electroburning process.[20]

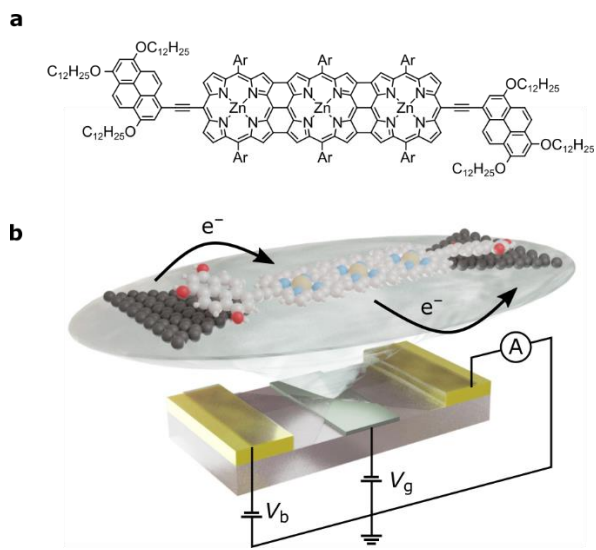

**Figure 1.** (**a**) The molecular structure of **FP3**: the edge-fused porphyrin trimer core is functionalized in the terminal *meso* positions by tridodecyloxypyrene groups for anchoring to graphene source and drain electrodes. Ar groups are solubilizing aryl groups, 3,5-*bis*(trihexylsilyl)phenyl. (**b**) Device architecture: nanometer-separated graphene source and drain electrodes are electroburnt from a graphene ribbon between two gold electrodes. The graphene is patterned into a bowtie shape, and a local gate electrode separated from the molecule by a thin layer of HfO$_2$ (grey) is used to shift the molecular energy levels. For clarity, the bulky side-groups are omitted.

**Extended Hubbard Model**

We have previously shown that the porphyrin monomer with the same, electron-rich, TDP anchor groups is commonly found in the *N–1* state (where *N* is the number of electrons on the molecule in the neutral state) upon adsorption onto *p*-doped graphene electrodes at zero gate voltage, $V_g = 0$.[26] **FP3** is more readily oxidized when compared to the monomer (first oxidation potentials are –0.07 V and 0.04 V for **FP3** and monomer respectively, both with respect to Fc|Fc$^+$, see SI). Thus, **FP3** is likely to be in an oxidized form upon physisorption onto the graphene electrodes, (in fact we show that it is oxidized to the dicationic *N–2* **FP3**$^{2+}$ state at $V_g = 0$, *vide infra*). We can therefore safely attribute the sequential tunneling regions that are observed in the experimental stability diagrams as corresponding to the transitions between different charge states of **FP3** as electrons tunnel into and from the highest occupied orbitals (of the neutral species).

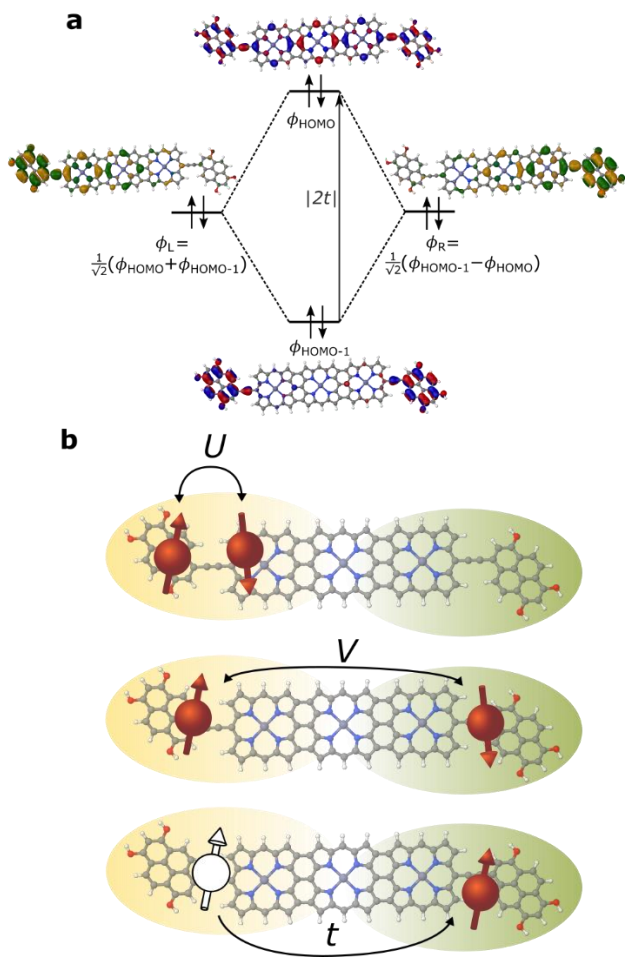

**Figure 2 (a)** MO diagram displaying the eigenbasis and site basis of the frontier orbitals of **FP3** in the $N$ state. Since the sum of the site orbitals yields the MO that is lower in energy the tunnel coupling, $t$, is negative. **(b)** Visualization of the Hubbard terms for **FP3**, $U$ and $V$ are potential energy terms due to on-site and inter-site repulsion, and $t$ is the kinetic energy term accounting for hopping between localized sites.

The (closely spaced) HOMO/HOMO–1 orbitals of **FP3**, the two orbitals emptied as the molecule is oxidized from $N$ to $N$–4 charge states (**FP3** to **FP3$^{4+}$**), are shown in Fig. 2a. The orbitals, by inspection, appear as an in-phase and out-of-phase (or bonding/anti-bonding) combination of 'site' orbitals that are primarily based on the electron-rich pyrene anchors (Fig. 2b). Thus, linear combinations of the delocalized HOMO/HOMO–1 can be taken to transform them into a localized 'left' and 'right' site orbital, $\phi_L$ and $\phi_R$ (Fig. 2a). By making this transformation, from an eigenbasis to site basis, the many-body electronic structure of **FP3** in the five oxidation states from $N \rightarrow N$–4 can be modelled using a two-site extended Hubbard dimer model in which the left site couples only to the left electrode and *vice versa*, and the two sites are coupled to each other.

The Hamiltonian of the full system is given by:

$$H = H_E + H_V + H_{HB} \tag{1}$$

where the left (L) and right (R) electrodes are fermionic reservoirs described by:

$$H_E = \sum_{l=L,R} \sum_{k_l,\sigma} \epsilon_{k_l} c^+_{k_l,\sigma} c_{k_l,\sigma} \tag{2}$$

that are coupled to **FP3** *via* the Hamiltonian:

$$H_V = \sum_{l=L,R} \sum_{k_l,\sigma} V_{k_l} a^+_{l,\sigma} c_{k_l,\sigma} + \text{H.c.} \tag{3}$$

The extended Hubbard Hamiltonian that describes the many-body electronic structure of **FP3** is given by:

$$H_{HB} = \sum_{i,\sigma} \epsilon_i n_{i,\sigma} + t \sum_{\sigma} (a^+_{L,\sigma} a_{R,\sigma} + a^+_{R,\sigma} a_{L,\sigma})$$
$$+ U \sum_i \left(n_{i,\uparrow} - \frac{1}{2}\right)\left(n_{i,\downarrow} - \frac{1}{2}\right) + V(n_{L,\uparrow} + n_{L,\downarrow} - 1)(n_{R,\uparrow} + n_{R,\downarrow} - 1) \tag{4}$$

where *t*, *U* and *V* are the inter-site tunnel coupling, on-site repulsion and inter-site repulsion, respectively (Fig. 2b). $a^+_{i,\sigma}$ and $a_{i,\sigma}$ are creation and annihilation operators for an electron of spin σ (= ↑ or ↓) in site *i* (= L or R). $n_{i,\sigma}$ are the number operators, $n_{i,\sigma} = a^+_{i,\sigma} a_{i,\sigma}$. Creation and annihilation operators for an electron of energy $\epsilon_{k_l}$ in the electrodes are given by $c^+_{k_l,\sigma}$ and $c_{k_l,\sigma}$ and $V_{k_l}$ is the coupling strength. We apply the wide-band approximation and take: $V_{k_l} = V_l$. This is related to the molecule-electrode coupling by: $\Gamma_l = 2\pi |V_l|^2 \rho_l$ under the assumption that the density of states in the leads, $\rho_l$, is constant.[27]

The energies of the molecular states, $\epsilon_i$ depend on the bias and gate voltages:

$$\epsilon_i = \epsilon_0 - \alpha_s V_b - \alpha_g V_g \tag{5}$$

where $\alpha_s$ and $\alpha_g$ are the coupling to the source and gate electrodes respectively.

For the two-site molecular system, which can accommodate up to four electrons, the eigenvectors of the Hubbard Hamiltonian are summarized in Table 1.[28] The 'vacuum' state corresponds to both HOMO and HOMO-1 being empty (*N*–4: **FP3**[4+]); the neutral molecule (*N* state) is when the HOMO–1 and HOMO are both filled. Therefore, there is a single electronic state for *N*–4 and *N* charge states. For each of *N*–1 and *N*–3 there are a pair of doubly (spin) degenerate states, separated in energy by 2|*t*|, denoted D⁻ and D⁺. Finally for *N*–2 there exist 6 states: a 3-fold degenerate triplet T, and three singlet states, analogous to a 2-orbital-2-electron treatment.[29] The nature of the singlet states is more complex than an open-shell/closed-shell description, as shown in Table 1.

The coefficients in Table 1, $c_+$ and $c_-$ are given by:

$$c_{+/-} = \frac{1}{2}\sqrt{1 \pm \frac{U-V}{2C}} \qquad (6)$$

where $C$ is given by:

$$C = \sqrt{\left(\frac{U-V}{2}\right)^2 + 4t^2} \qquad (7)$$

giving some of the *N–2* singlet states a mix of open-shell/closed shell character that depends on the size of the on-site and inter-site repulsion, and inter-site hopping.

**Table 1.** The 16 energy eigenvectors of the Hubbard Hamiltonian for charge states *N–4* to *N* and their designation, under the assumption of equal site energies. $\sigma = (\uparrow,\downarrow)$, $\Phi_A = |\uparrow\downarrow,0\rangle$, $\Phi_B = |0,\uparrow\downarrow\rangle$, $\Phi_C = |\uparrow,\downarrow\rangle$ and $\Phi_D = |\downarrow,\uparrow\rangle$. The coefficients $c_+$ and $c_-$ depend on the values of *t*, *U*, and *V*, as described in the text.

| Charge state | Eigenstates of $H_{HB}$ | State (degeneracy) |
|---|---|---|
| *N – 4* | $|0,0\rangle$ | $S^{N-4}$ (1) |
| *N – 3* | $(|\sigma,0\rangle + |0,\sigma\rangle)/\sqrt{2}$ | $D_{+,\sigma}^{N-3}$ (2) |
|  | $(|\sigma,0\rangle - |0,\sigma\rangle)/\sqrt{2}$ | $D_{-,\sigma}^{N-3}$ (2) |
| *N – 2* | $c_-(\Phi_A + \Phi_B) - c_+(\Phi_C - \Phi_D)$ | $S_-^{N-2}$, (1) |
|  | $(|\uparrow,\downarrow\rangle + |\downarrow,\uparrow\rangle)/\sqrt{2}$, $|\uparrow,\uparrow\rangle$, $|\downarrow,\downarrow\rangle$ | $T_0^{N-2}$, $T_1^{N-2}$ $T_{-1}^{N-2}$ (3) |
|  | $(|\uparrow\downarrow,0\rangle - |0,\uparrow\downarrow\rangle)/\sqrt{2}$ | $S_{CS}^{N-2}$, (1) |
|  | $c_+(\Phi_A + \Phi_B) + c_-(\Phi_C + \Phi_D)$ | $S_+^{N-2}$, (1) |
| *N – 1* | $(|\uparrow\downarrow,\sigma\rangle + |\sigma,\uparrow\downarrow\rangle)/\sqrt{2}$ | $D_{+,\sigma}^{N-1}$, (2) |
|  | $(|\uparrow\downarrow,\sigma\rangle - |\sigma,\uparrow\downarrow\rangle)/\sqrt{2}$ | $D_{-,\sigma}^{N-1}$, (2) |
| *N* | $|\uparrow\downarrow,\uparrow\downarrow\rangle$ | $S^N$, (1) |

For transport through a number of electronic states, the current through the molecular junction can be compactly calculated using a rate-equation-type framework by first constructing the (in this case 16 x 16) rate-equation matrix, $\boldsymbol{W}$.[28,30] Taking the steady state approximation, $d\boldsymbol{P}/dt = \boldsymbol{WP} = 0$, the stationary occupation probabilities of the 16 electronic states $\boldsymbol{P_0}$ are the null space of $\boldsymbol{W}$, normalized such that the elements of $\boldsymbol{P_0}$ are non-negative and sum to 1.[30] The total current is then calculated by considering the tunneling processes at either electrode. The elements of $\boldsymbol{W}$ are comprised of the electron-transfer rates between states $j$ and $k$, $\gamma_l^{j \to k}$.

The electron-transfer rates are given by:

$$\gamma^l_{j,N \to k,N+1} = |D_{jk}|^2 \frac{\Gamma_l}{\hbar} \int f_l(\epsilon) k(\epsilon)^{j \to k} d\epsilon \qquad (8)$$

$$\gamma^l_{k,N+1 \to j,N} = |D_{kj}|^2 \frac{\Gamma_l}{\hbar} \int (1 - f_l(\epsilon)) k(\epsilon)^{k \to j} d\epsilon \qquad (9)$$

for reduction and oxidation respectively at electrode $l(=L/R$ for left/right electrode) . $\Gamma_l$ is the molecule-electrode coupling, $f_l(\epsilon)$ is the Fermi-Dirac distribution of electron energies in electrode $l$. The electron-transfer rate constants, $k(\epsilon)^{j \to k}$, are Dirac delta functions centered at the chemical potential of the transition from $j$ to $k$. As we will show later, these functions can be replaced with energy-dependent rate constants that also account for electron-vibrational coupling accompanying electron-transfer. $D_{jk}$ is the overlap integral $\langle \phi_k | a^+_{i,\sigma} | \phi_j \rangle$, also known as the Dyson orbital coefficient. Inclusion of these coefficients precludes the need to include statistical factors based on degeneracies into the rate-equation matrix. Furthermore, they automatically encode the selection rules for electron transfer: $\Delta S = \pm 1/2$ and $\Delta m_S = \pm 1/2$. For instance, the transitions from the $D^{N-3}_{+,\uparrow}$ state to the $T^{N-2}_0$, $T^{N-2}_1$ $T^{N-2}_{-1}$ states result in Dyson coefficients of 1/2, 1/√2 and 0, respectively. That is, if the molecule is in a $D^{N-3}_{+,\uparrow}$ state, it has a single, spin-up electron (m_S = ½), and so one additional electron cannot hop in to create the state $T^{N-2}_{1,\downarrow}$ which has two spin-down electrons (m_S = −1) so $\gamma^l_{D+,N-3 \to T_{-1},N-2} = 0$.

**Experimental Charge Stability Diagrams**

Stability diagrams of **FP3** devices are given in Fig. 3 (device **A**) and SI (device **B** and **C**). Multiple sequential tunneling regions are observed, as expected, reflecting the redox activity of the edge-fused trimer with respect to the monomer.[7] A common feature of the charge stability diagrams of devices **A**–**C** is the presence of a larger Coulomb diamond centered around $V_g = 0$, flanked by smaller diamonds with addition energies ranging from 0.14 – 0.30 eV. The addition energy, $E_{add}$, is the energy required to add an extra electron to a molecule in the device, and can be read directly from a stability diagram as the width of the corresponding Coulomb diamond (scaled by the gate coupling, $\alpha_G$).[21] The total energies eigenvalues of the Hubbard Hamiltonian can be translated into analytical expressions for addition energies of the *N–1, N-2* and *N–3* charge states by taking the energy of the ground state of each charge state. In the limit of $U, V \gg t$, the addition energies are *V–t* for the odd diamonds *N–1* and *N–3,* and *U* for *N–2*.

By considering the experimental addition energies of device **A** in the extended Hubbard framework, we are able to determine the electron-electron repulsion terms $U \approx 0.5$ eV and $V \approx 0.14$ eV; from the rectification behavior and DFT calculations (discussed below), we infer below that $U, V \gg t$. We know that $|t|$ must be non-zero (for transport to occur) and negative (due to the spatial properties of the orbitals, see Fig. 2). The value of $t = -0.01$ eV is in agreement with our later experimental observations. The calculated stability diagram, using coupling to the electrodes determined from the slopes of the experimental Coulomb diamonds, is shown in Fig. 3b. Not only does the extended Hubbard model predict the positions of the edges in the stability diagram, it is also simultaneously consistent with other experimental observations, i.e. gate-dependent rectification ratios and high-bias excited states, as outlined in the following paragraphs.

For transport through a single spin-degenerate level the rectification ratio varies between 1/2 and 2 depending on the asymmetry of the molecule-electrode coupling.[26] The *N–4/N–3* transition (at $V_g = -66$ V in Fig. 4a) exhibits a rectification ratio of nearly exactly 4 (see Fig. 4c), a clear deviation from what is observed for a single spin-degenerate level. That rectification ratio is also present in the corresponding *IV* traces of device **B** and **C**.

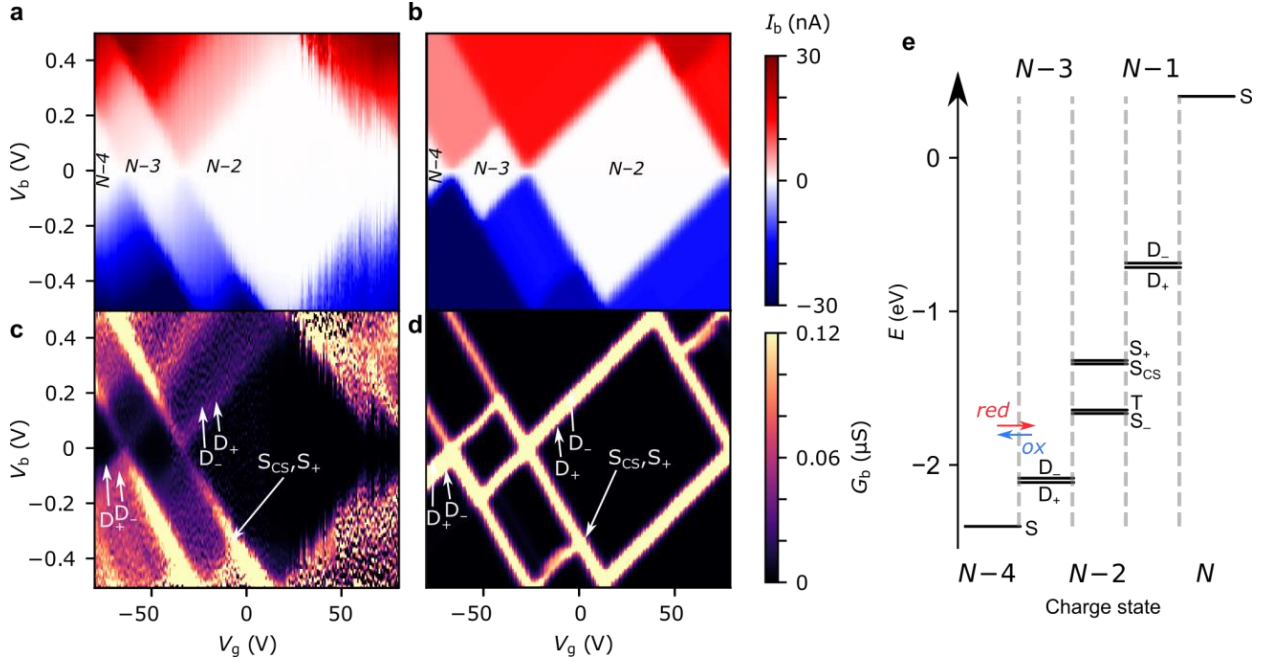

**Figure 3.** Experimental stability diagrams of device **A**, (**a**) current and (**c**) derived conductance, measured at 77 K; the Coulomb diamonds are assigned with their charge states. (**b**) and (**d**) are current and conductance stability diagrams, respectively, calculated using the Hubbard model, with parameters $U = 0.50$ eV, $V = 0.14$ eV and $t = -0.01$ eV. The couplings to the source and gate electrodes are $\alpha_S = 0.37$, and $\alpha_G = 4.8 \times 10^{-3}$, taken from the experimental data. The molecule-electrode couplings are taken from the *IV* fit in Fig. 3a. (**e**) The energy eigenvalues of the 16 Fock states of **FP3** involved in electron transfer calculated for device **A** using the Hubbard model at $V_G = -80$ V, the lowest experimental gate voltage.

Due to the asymmetry in the molecule-electrode couplings (present in all devices considered here) the rectification ratio of the resonant *IV* trace can be used to directly infer the number of electronic states within the bias window.[31, 32] For device **A**, $\Gamma_R \gg \Gamma_L$, and so a ratio $I_b(-V_b):I_b(+V_b)$ of 4:1 indicates the rate of tunneling onto the molecule is four times greater than tunneling off. From this we can infer that there are four available states for reduction of the *N–4* state, but only a single state that can be accessed from the oxidation of the *N–3* state. The observed rectification in the experimental data is inherently captured by the Hubbard model (Fig. 4a and c). The energy spacing between the $D_+^{N-3}$ and $D_-^{N-3}$ levels is only $2|t| = 20$ meV, and therefore both levels are found within the bias window at above roughly 50 mV. Once both the low-lying doublets $D_+^{N-3}$, and $D_-^{N-3}$ are within the bias window, there are four *N–3* states that $S^{N-4}$ can be reduced to when an electron hops onto the **FP3**$^{4+}$, but each state can only be oxidized back to $S^{N-4}$, see Fig. 3e. At 77 K, the transition from 1:2 to 1:4 rectification ratios as $D_-^{N-3}$ enters the bias window is significantly broadened due to the lifetime-broadening, the Fermi functions in the leads, but mainly due to the energy-dependence of the hopping rates that results from electron-vibrational coupling (not accounted for in the Hubbard model). Therefore, the excited state transition $S^{N-4} \leftrightarrow D_-^{N-3}$ is not visible as a separate parallel line intersecting the *N–3* Coulomb diamond, and, this observation is consistent with an estimated value of *t* that is of the same order as $k_BT$ ($8\ meV$).

For the *N–3/N–2* transition (at $V_g$ = –34 V), the experimental rectification ratios are around 1 (see Fig. 4b and c). This is again in agreement with the Hubbard model. From Fig. 3e, this ratio arises because charge transport at higher bias occurs between four doublets of the *N–3* charge state and the ground-state singlet and the low-lying triplet of the *N–2* charge state (the singlet-triplet gap is only approximately ~1 meV for the values of parameters used in the Hubbard model). The remaining excited singlet states are visible in Fig. 3 as the excited state line at higher bias. The situation becomes slightly more nuanced as the probability of each transition is scaled by a relevant Dyson orbital coefficient. The rectification behavior can be understood for the full set of transitions from Fig. 4d, the off-diagonal elements connecting two charge states represent the rate of transfer between those two states on resonance. By inspection of the upper left corner we can see that the rate of *N–4* → *N–3* is 2.0 whereas it is 0.5 for *N–3* → *N–4*, giving a ratio of $I_b(-V_b):I_b(+V_b)$ of 4:1. For *N–3/N–2* it is 1.0 for either direction. These values also reflect the relative magnitudes of current expected between the *N–4/N–3* and *N–3/N–2* transitions for a Hubbard dimer, as is observed experimentally.

**Electron-Vibrational Coupling**

Due to their relatively small size, molecular systems undergo significant geometric changes upon charging when compared to lithographically defined structures, and as such vibration coupling to sequential electron transport is significant for these systems. **FP3** has $3N_{atom}-6 = 3345$ vibrational normal modes that span the energies between a few meV for out-of-plane bending motions, through several hundreds of meV for C-C bond stretches and up to 400 meV for C-H stretches, as is typical of a large π-conjugated molecule. Electron-vibration coupling to these modes causes low-bias suppression of tunneling current, and by omitting them, *IV* traces calculated from the Hubbard model significantly overestimate the current at low bias, as can be seen in Fig. 4a and Fig. 4b. The absence of electron-vibration coupling in the Hubbard model also accounts for the lack of asymmetry in the sequential tunneling regions with respect to gate voltage that is visible in the experimental stability diagrams.[26] In order to reproduce absolute values of the current and therefore reinforce the fact that the two-site Hubbard model is applicable, we incorporate electron-vibration coupling into the electron-transfer rate constants, $k^{i \to j}$, by replacing the Dirac delta functions centered on the chemical potential of the transition from *i* to *j*.

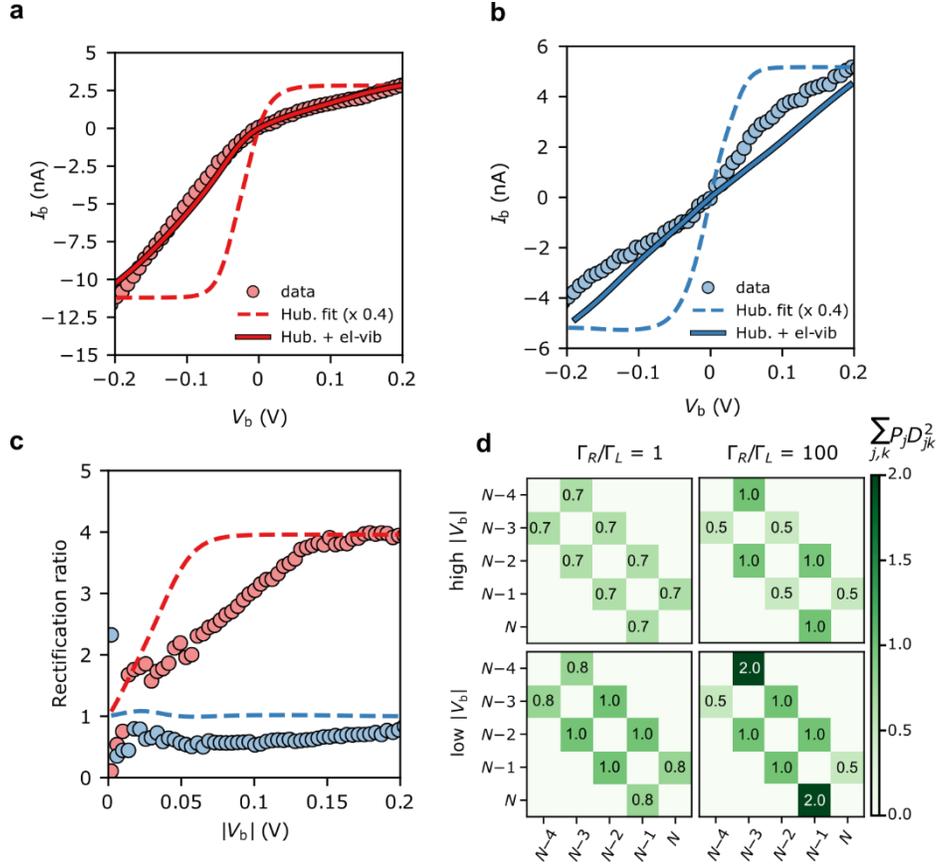

**Figure 4.** Device **A** *IV* traces on (a) the *N–4/N–3* resonance ($V_G = -66$ V), and (b) the *N–3/N–2* resonance ($V_G = -34$ V). The experimental data are plotted alongside *IV* traces taken from the Hubbard stability diagrams in Fig. 3, and the Hubbard model plus electron-vibration coupling included in the electron-transfer rates. Electron-vibration fitting parameters for *N–4/N–3*: $\Gamma_L = 41\ \mu eV, \Gamma_R = 14\ meV, \lambda_o = 70\ meV$; for *N–3/N–2*, $\lambda_o = 120\ meV$. (c) The rectification behavior, $I_b(-V_b): I_b(+V_b)$, of the *N–4/N–3* (red) and *N–3/N–2* (blue) transitions are given for the experimental values (circles) and the Hubbard model (dashed lines). The measurements were taken at a device temperature of 77 K. (**d**) The emergence of the rectification behavior of the Hubbard model under asymmetric molecule-electrode coupling. Each element (*j*,*k*) represents the sum of the electron transfer rates from charge state *j* to *k*. 'High' $|V_b|$ means above $2|t|/\alpha_S$, when both *N–3* doublets are within the bias window. An improved fit to *N–3/N–2* (maintaining the rectification ratio) can be achieved by going beyond the wide-band gap approximation, see SI.

The method we use follows previous work,[7, 27] and is described in more detail in the SI. In short, a spectral density is constructed that accounts for contributions to the rates of electron transfer from the inner sphere (i.e. distortion of the molecule along normal modes of vibration upon charging) and from the outer sphere (i.e. distortion of the local molecular environment, predominantly the substrate). For the *N–4/N–3* transition, the electron-transfer rates for reduction: $k^{S \to D+}$ and $k^{S \to D-}$ (and similarly for oxidation, $k^{D+ \to S}$ and $k^{D- \to S}$) are assumed to be the same except for the offset in energy by spacing between the doublets, $|2t|$. The experimental *N–4/N–3* *IV* traces are then fitted using three parameters, $\lambda_o$, $\Gamma_S$, and $\Gamma_D$, to reproduce the experimental data (Fig 4a.). The fits to the *N–4/N–3* transitions for device **B** and **C** are given in the SI.

The *N–3/N–2* resonant *IV* curve can be fitted following the same method. The geometry of the *N–2* state (**FP3$^{2+}$**) is optimized in the singlet or triplet ground state to calculate $k^{D+ \to S-}$ and $k^{D+ \to T}$. As with the *N–4/N–3* transition, we assume the geometry of the doublets, $D_+^{N-3}$ and $D_-^{N-3}$ are the same. The molecule-electrode couplings from the *N–4/N–3* fit are used and therefore $\lambda_o$ is the only free parameter. Fig. 4 shows the inclusion of electron-vibration coupling converts the Hubbard *IV*s, which give the required rectification ratios, into good fits to the experimental data.

**DFT Calculations**

The addition energies for **A**, **B**, and **C** are given in Fig. 5a; the devices follow the same trend with only slight variations in the values of *U*, *V*, and *t* needed to be selected for each device. The values of these parameters extracted from the experimental charge stability diagrams, which are seemingly intrinsic to the molecular structure, can be compared to those calculated using DFT. Electron-electron repulsion terms are the Coulomb integrals: $U = \left\langle \phi_L \phi_L \left| \frac{1}{4\pi\epsilon_0\epsilon_r r_{12}} \right| \phi_L \phi_L \right\rangle$ and $V = \left\langle \phi_L \phi_L \left| \frac{1}{4\pi\epsilon_0\epsilon_r r_{12}} \right| \phi_R \phi_R \right\rangle$. For optimized gas-phase geometries ($\epsilon_r = 1$) the values are $U_{DFT} = 2.62$ eV and $V_{DFT} = 1.0$ eV (Fig. 5c). The kinetic energy, *t*, is obtained from DFT calculations as half the difference between the HOMO/HOMO–1 (see Fig. 2a), which yields $t_{DFT} = -0.13$ eV. Values very similar to those obtained experimentally can be obtained by introducing an effective $\epsilon_r$ that accounts for the dielectric environment. If we set $\epsilon_r$ to 5.5 (a value comparable other π-conjugated organic molecules[33]) we obtain $U'_{DFT} = 0.47$ eV, and $V'_{DFT} = 0.18$ eV. This effective dielectric constant can account for intramolecular charge screening as well as polarization of the oxide substrate and the graphene electrodes.[34] The kinetic energy term *t* does not scale linearly with $\epsilon_r$, however it can still be altered by the electrostatic influence of the substrate and geometric distortions. The binding energies of the anchor groups to the graphene are estimated to be several eV,[35] and therefore, for each molecular junction, where the exact atomic structure of the electrodes are unknown, the molecule can readily adopt a unique conformation to maximize binding. As an example of one of many possible low-energy distortions of the molecular geometry, $t$ has a strong dependence on the dihedral angle between the porphyrin trimer and the anchor groups (see SI). In addition, this coordinate modifies the **FP3$^{2+}$** singlet-triplet energy spacing (which are calculated using the Hubbard model to be 1.2 meV, 1.1 meV, and 30 meV for devices **A**, **B** and **C,** respectively). Therefore, the device-to-device variation observed is most likely due to both differences in molecular conformation, and the unique local dielectric environment for each device. This further highlights the requirement that if highly reproducible single-molecule device characteristics are desired, precise control over the molecular environment is necessary.

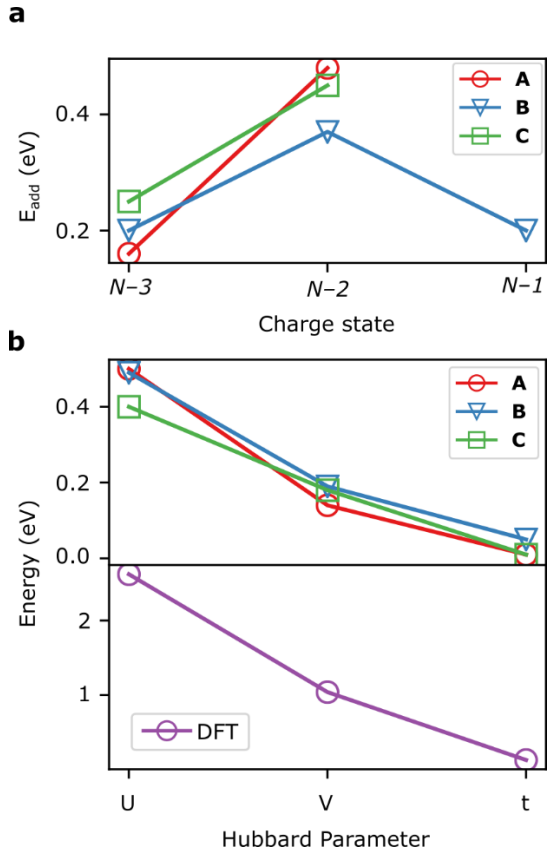

**Figure 5.** (**a**) Addition energies from experimental stability diagrams for devices **A**–**C**. (**b**) The Hubbard parameters for the devices that reproduce these addition energies (upper panel) along with DFT calculations of these values (lower panel).

**Conclusions**

In conclusion, we report on the sequential transport behavior of an edge-fused porphyrin trimer in a single-molecule junction at 77 K. The large, conjugated molecular structure and weak molecule-electrode coupling lead to multiple sequential tunneling regions that are experimentally accessible. This allows us to study the many-body electronic structure of this system in various charge states, and understand the resulting transport properties of the junction. Due to the spatial distribution of the orbitals involved in transport, **FP3** can be modelled as two-site Hubbard dimer. Uniquely amongst related two-site molecules,[36, 37] the molecule is fully conjugated between the two sites, apparently negating any voltage drop across the molecule. The Hubbard framework reproduces key features of the experimental stability diagrams that pertain to many-body electron-electron interactions, i.e. the addition energies, rectification ratios (which indicate the presence of excited states involved in electron transfer), and high bias excited states. A quantitative reproduction of the experimental *IV* curves requires integrating electron-vibrational interactions into the Hubbard model. These experiments may guide future explorations of the role of electron correlations in charge transport through extended aromatic systems.

**Methods**

**Device Fabrication**

Device fabrication followed previously reported procedures.[7, 20] Devices were fabricated on n-doped silicon wafers with 300 nm of thermally grown $SiO_2$. For device **A** and device **B** the underlying doped silicon was used as a global gate for all devices on each chip. For device **C** a local gate electrode was patterned. The local gates were fabricated by optical lithography and e-beam evaporation of titanium (10 nm) and gold (30 nm). A dielectric layer of $HfO_2$ (10 nm) was subsequently deposited by atomic layer deposition. Source and drain contact pads were patterned onto the $SiO_2$ (device **A** and **B**) or $HfO_2$ (device **C**) by optical lithography and e-beam evaporation of titanium and gold (10 nm / 60 nm for devices **A** and **B**, 10 nm / 60 nm for device **C**).

CVD-grown monolayer graphene was transferred onto the devices by Graphenea. The graphene was patterned into bow-tie shapes with a width of approximately 100 nm at the narrowest point. First the devices were spin-coated with the negative tone resist ma-N 2403 and patterned using e-beam lithography with a dose of 120 $\mu C\ cm^{-2}$ and an accelerating voltage of 50 kV. The pattern was developed with ma-D 525 to remove the unexposed resist, and unprotected regions of graphene were etched by $O_2$ plasma. The developed resist was removed with an NMP-based remover REM660 to give the bowtie-shaped graphene. Finally, the patterned graphene was formed into a nanometer-spaced graphene tunnel junctions, graphene nano-gaps, by feedback-controlled electroburning[24, 25] with a threshold resistance of 600 MΩ. The IV curves after electroburning were fitted with the Simmons model to estimate the spacing between the graphene source and drain electrodes to be around 1.5 nm.

The gate-dependence of the source-drain current was measured at room temperature before deposition of the molecular solution on an automated probe station. The molecules were deposited onto the graphene nano-gaps from a 2 μM toluene solution, and the devices were measured again.

**Electrical Measurements**

Device **A** and device **B** were wire-bonded into a chip carrier and measured in a dip-stick setup at 77 K. The dip-stick was evacuated and immersed in a dewar of liquid nitrogen. A HP33120A function generator was used to apply the source-drain voltage. The gate voltage was applied by a Keithley 2450 SourceMeter. A Stanford Research Systems SR570 low-noise current amplifier was used to measure the source-drain current, and the data collected by a National Instruments BNC-2090A DAQ. Device **C** was measured in an Oxford Instruments 4K Pucktester. An Adwin Gold II data acquisition system was used to apply the source-drain and gate voltages. An SR570 was used to measure the current which was collected by the Adwin Gold II.

ASSOCIATED CONTENT

**Supporting Information**.

The following files are available free of charge.

Synthetic procedures and characterization, supporting device data, calculation of electron-phonon coupling constants, supporting DFT calculations. (PDF)

ACKNOWLEDGMENT

This work was supported by the EPSRC (grants EP/N017188/1 and EP/R029229/1). J.A.M. acknowledges funding from the Royal Academy of Engineering and a UKRI Future Leaders Fellowship, Grant No. MR/S032541/1. JKS and GCS thank the Department of Energy, Office of Basic Energy Sciences, grant DE-SC0000989, for theory work.